\documentclass{article}

\usepackage{arxiv}

\usepackage[utf8]{inputenc} % allow utf-8 input
\usepackage[T1]{fontenc}    % use 8-bit T1 fonts
\usepackage{hyperref}       % hyperlinks
\usepackage{url}            % simple URL typesetting
\usepackage{booktabs}       % professional-quality tables
\usepackage{amsfonts}       % blackboard math symbols
\usepackage{nicefrac}       % compact symbols for 1/2, etc.
\usepackage{microtype}      % microtypography
\usepackage{lipsum}		% Can be removed after putting your text content
\usepackage{graphicx}

\usepackage{csquotes}

\usepackage{graphicx}
\usepackage{subcaption}
\usepackage{todonotes}
\usepackage[mathscr]{euscript}  % required for math
\usepackage{amsfonts}           % required for math

\title{Drawing with AI - Exploring Collaborative Inking Experiences based on Mid-air Pointing and Reinforcement Learning}

\author{ \\Franziska Geiger, Michelle Martin, Monika Pichlmair, Ilhan Aslan, Hannes Ritschel,\\ Björn Bittner, and Elisabeth André \\
Human-Centered Multimedia Lab\\
Augsburg University\\
Germany\\
\texttt{name.lastname@student.uni-augsburg.de, lastname@hcm-lab.de} 
}

\begin{document}
\maketitle

\begin{abstract}
    Digitalization is changing the nature of tools and materials, which are used in artistic practices in professional and non-professional settings. 
    For example, today it is common that even children express their ideas and explore their creativity by drawing on tablets as digital canvases. 
    While there are many software-based tools, which resemble traditional tools, such as various forms of virtual brushes, erasers, etc. in contrast to traditional materials there is potential in augmenting software-based  tools and digital canvases with artificial intelligence. 
    Curious about how it would feel to interact with a digital canvas, which would be in contrast to a traditional canvas dynamic, responsive, and potentially able to continuously adapt to its user's input, we developed a drawing application and conducted a qualitative study with 14 users. 
    In this paper, we describe details of our design process, which lead up to using a k-armed bandit as a simple form of reinforcement learning and a LeapMotion sensor to allow people from all walks of like, old and young to draw on pervasive displays, small and large, positioned near or far. 		
	
\end{abstract}

	\begin{figure}
	    \centering
		\includegraphics[width=\textwidth]{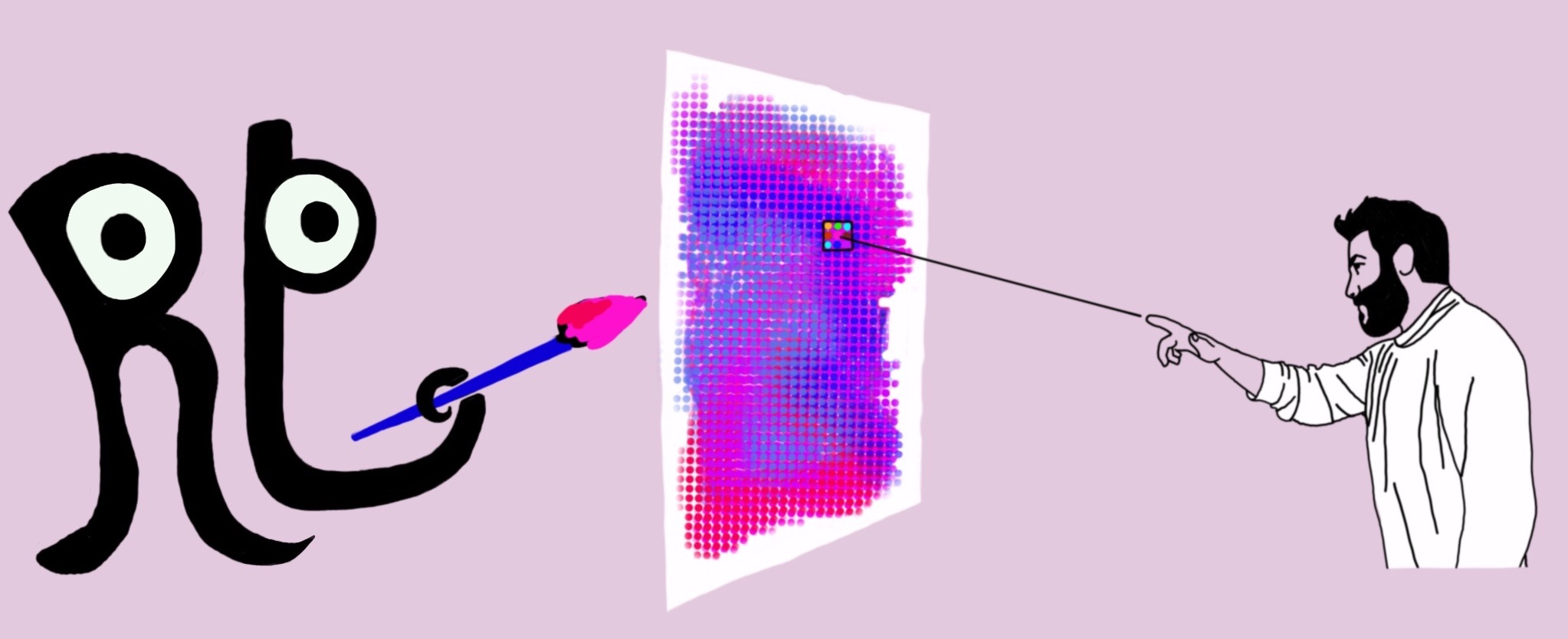}
		\caption{Drawing on a pervasive display (i..e, a digital canvas) that tries to learn its user's color preferences and inks ``pixels'' on the canvas accordingly. }
		\label{fig:teaser}
	\end{figure}

\section{Introduction}

The number of application domains that are being influenced by machine learning is expanding day by day, and assistance for personal health (e.g., \cite{flutura2018drinkwatch})
wellbeing  (\cite{rathmann2020towards}) are only two examples. Creative practices in the fields of design and art do not seem to be excluded from this progress and transformation in the field of software development. In a recent article, Seidel and colleagues \cite{seidel2018autonomous} discuss the growing importance of autonomous tools, which are capable to produce great outcomes and that therefore designers need to adopt the usage of such tools and adapt their work processes. They argue: \begin{quote} It is important for designers to understand how their own mental models interact with mental models embedded in the logic of autonomous tools  \end{quote}

It may be true that the way we, as designers think about our digital tools has to change, with us increasingly perceiving autonomous tools as agents to collaborate with, instead of considering them simply as extensions of our bodily skills. And, mental models of traditional drawing materials, such as canvases, pens, and colors are indeed being challenged by more and more people growing up using digital tools enabled by off-the-shelf devices, such as tablets. 
But, how we draw on screens has been inspired for a long time by how we draw on paper; and, users' imagination is often still constrained by the materiality of the original physical drawing materials and the original form of contact-based pen and paper interaction. While  ever new pervasive displays, such as architectural sized screens and novel forms of interactions using a slingshot metaphor (e.g.\cite{fischer2012urban}) challenge people's mental models, it seems unclear how non-professionals would feel about interacting with a pervasive display as a digital drawing canvas, which is autonomous to some degree.  

In order to explore this notion of interacting with an intelligent digital canvas, we developed a drawing application based on reinforcement learning. Since mid-air pointing seems to be an engaging, ``natural'', and flexible way to draw on pervasive displays (e.g., ~\cite{chen2017identifying, Alt:2018:STM:3205873.3205887}) we decided to integrate it as an alternative interaction technique for users. Before we describe the design of the drawing application in detail and report on a qualitative user study with a heterogeneous group of users we provide the necessary background in reinforcement learning and related work in mid-air interaction in the next section.

\section{Background and Related Work}

\subsection{Reinforcement Learning}
In general, machine learning is split up into three main categories: supervised, unsupervised, and reinforcement learning.
While the first is learning from labeled examples provided by an expert supervisor, the second extracts structures from unlabeled data.
In contrast, reinforcement learning \cite{sutton2018reinforcement} is used for decision-making based on trial and error: the learning agent, which is situated in an \emph{environment}, incrementally explores which \emph{actions} to take in different situations, called \emph{states}.
In each time step \(t\) the learning agent determines its current state \(S_t \in \mathscr{S}\) from the set of states \(\mathscr{S}\), manipulates the environment by executing action \(A_t \in \mathscr{A}(S_t)\) from the set of actions \(\mathscr{A}(S_t)\) available in the current state and receives a scalar \(R_{t+1} \in \mathbb{R}\) as feedback, the so-called \emph{reward}.
The magnitude of \(R_{t+1}\) serves as an indicator of how ``good'' or ``bad'' action \(A_t\) performs in state \(S_t\).
By iteratively exploring actions in different states and receiving the corresponding reward signals, the agent can improve its own behavior, called \emph{policy}, over time.

Reinforcement learning does not need an expert supervisor or (un)labeled data upfront.
The agent acts on its own and finds a solution over time, solely based on the reward it receives depending on the current state and executed action.
While there are different approaches for optimizing the agent's behavior and extracting the optimal policy, we focus on value-based algorithms with discrete state and action space.
In each learning step the learning agent updates the value \(Q_t\) of the current state-action pair.
In traditional Q-Learning, this is done by calculating \( Q(S_t, A_t) = Q(S_t, A_t) + \alpha \left[ R_{t+1} + \gamma max_a Q(S_{t+1}, a) - Q(S_t, A_t) \right] \).
Thereby, the \emph{learning rate} \( \alpha \in [0;1] \) controls the speed of learning and the \emph{discount factor} \(\gamma \in [0;1] \) reduces future rewards in order to encourage the agent to solve the problem most efficiently.
A reinforcement learning agent's behavior is \emph{greedy}, i.e. it aims to maximize received rewards by primarily selecting the action with the largest \(Q\)-value in each state.

In stateless environments, where the agent does not need to distinguish different situations, \(k\)-armed bandit problems \cite{sutton2018reinforcement} are a simple form of reinforcement learning.
The agent only distinguishes \(k\) different actions \( \mathscr{A} \) with \( |\mathscr{A}| = k \), but no states.
Thus, the agent learns which of the \(k\) actions is the best by repeatedly executing actions, receiving the rewards and calculating \(Q(a)\) for each \(a \in \mathscr{A} \).

Being a completely autonomous learning approach makes reinforcement learning interesting for tasks where the agent's optimal behavior can hardly be defined upfront because of lacking domain knowledge or complexity.
Therefore, this machine learning approach is also used in the context of HCI (human-computer interaction) and HRI (human-robot interaction), e.g. to optimize dialog systems \cite{DBLP:series/tanlp/RieserL11}, to explore optimal robot behavior for story telling \cite{DBLP:conf/ro-man/RitschelBA17}, presentation of jokes \cite{DBLP:conf/icmi/WeberRALA18, ritschel2019irony} or argumentation \cite{ritschel2018drink}.

\subsection{Mid-air Interaction and User Experience}
\label{ssec:LeapMotion}
Previous work has extensively explored mid-air input as a modality for human-computer interaction and designed its use for contexts, such as cars \cite{Aslan:2015:AUI}, clean rooms in industrial settings \cite{Aslan:2016:DEM:2997043.2832919}, and for retail situations \cite{aslan2015sharing}, but also for pervasive displays~\cite{chen2017identifying, Alt:2018:STM:3205873.3205887}. Aslan et al. \cite{aslan2017pre} showed that interfaces reacting to the movement of a user's hand before any touch event happens can appear more attractive to the user.
Especially changes that were perceived as "life like", like a target changing color or form, seemed pleasant.
The authors figured out that proxemic touch targets might not be very beneficial for interfaces used in professional environments, e.g. a work setting where efficiency is more important than aesthetics. It could, however, be very useful for novice users and interaction settings that benefit from a slower pace like creative tasks, and also in combination with a smart pen~\cite{Aslan:2018:EUE:3282894.3282906}.

When sensing the mid-air space, the LeapMotion controller is often used. The LeapMotion is a device for precise hand and gesture recognition, containing two infrared cameras and three infrared LEDs. The controller projects a pattern of points within a 135 degree~field around the device and analyses the data captured by the infrared cameras~\cite{leapHP}.
The LeapMotion API processes the collected data and offers information about the detected hands, fingers and tools in 3D space~\cite{leitao2015analysis}. 

While the LeapMotion achieves very high accuracy in static situations~\cite{weichert2013analysis}, the accuracy in dynamic situations can be lower due to, for example, hand tremors~\cite{valentini2017accuracy}, which must be taken into account (e.g., by making targets big enough).

\section{Design}

New technologies and interaction methods will only be established in real world for interfaces if they are designed highly intuitive. 
To ensure users would not be overwhelmed by the complexity of the system, we aimed at creating a simple adaptive user interface. 
It should be usable for people without knowledge or interest in computer science and technology in general. 
Even people who have never been confronted with AI or tracking devices should be able to interact with it intuitively. These requirements should be fulfilled to archive a positive user experience and provide a smooth handling of our application.

Out of this we decided to used a simple matrix grid as base user interface. By using the LeapMotion controller and gestures, the user is then able to interact with the system and change field colors on the grid. 
The exact structure of the user interface is described in section \ref{sec:UI}.

\subsection{Ideation - Brainstorming} 

During our brainstorming phase we explored various designs. Our first idea was a maze-game where the user should be led through an invisible 3D-grid, following a predefined path. The system should thereby learn to guide the user through the path by showing customized colors on the grid.
As starting position an empty 2D-grid was shown on screen. 
Once the user pointed at a panel, the panels in a von Neumann neighborhood around the pointed-on panel changed their color. The von Neumann neighborhood describes the four directly adjoining fields of the pointed-on paneln. 
For choosing these colors, a simple reinforecement learning algorithm, namely a so called Q-learner was responsible. 
Its goal was to lead the user's hand along a predefined invisible path, e.g. encouraging movements up or down, farther away from the screen and so on. 
We explored whether the Q-learner could learn to optimize it's color output even if we use only a small number of actions. 
Additionally, we wanted to investigate if the user would be able to interpret the systems behavior over time. 
This should result in an interaction loop where both parts would try to understand the given information and adapt their reactions according to what they perceived. 

Based on initial tests with this maze-game we identified several insights: 
\begin{itemize}
    \itemsep0pt
    \item The Q-learner does not learn fast enough for this application. 
        At the beginning of the training, it receives primarily negative feedback because the users tend to point randomly across the screen. 
        As a result, the user does not notice any learning progress and feels discouraged.
    \item The user tends to be confused by the lack of information presented. 
        Even simple static information like the previously pointed-on panels keeping their color would be helpful here. 
        Since we limited the number of displayed colors to three, showing more colored panels over time would also give the user a sense of progress. % mhaisma: again, need that image.
    \item Users tend to ignore the given task - trying to find a way through a 3D-maze - after a while and start playing around instead.
        This also lead to a lot of unwanted negative feedback for the Q-learner.
    \item Users reacted to different kinds of stimulation. 
        Some mainly moved towards the brightest color shown, others preferred the color with the greatest difference to the other colors in the neighborhood or were looking for different color-combinations. 
\end{itemize}

Based on these difficulties, we decided to design a second application, focused on drawing instead of path-finding.
We kept the grid and the colored neighborhood, but instead of trying to lead the user, we wanted to encourage them to move freely. 
They can permanently change the color of any panel in the colored neighborhood by pointing at it.
Additionally the opacity of the drawing color changes depending on the relative position of the finger to the tracking sensor. 
All colors within the neighborhood are chosen by a simpler reinforcement learning algorithm. The detailed implementation is explained in section \ref{sec:implementation}. 
In comparison to the first idea, the approach adapts quickly to the user's current preferences, provides faster feedback and a stronger sense of control. 
This ``game'' does not have an explicit goal, it is only used as a painting application. 
Instead, it encourages free movements and enables the user to fully focus on the drawing experience. Out of this focus we assume to gain better and clearer user study results. 

\begin{figure}[h]
    \centering
    \includegraphics[width=0.6\textwidth]{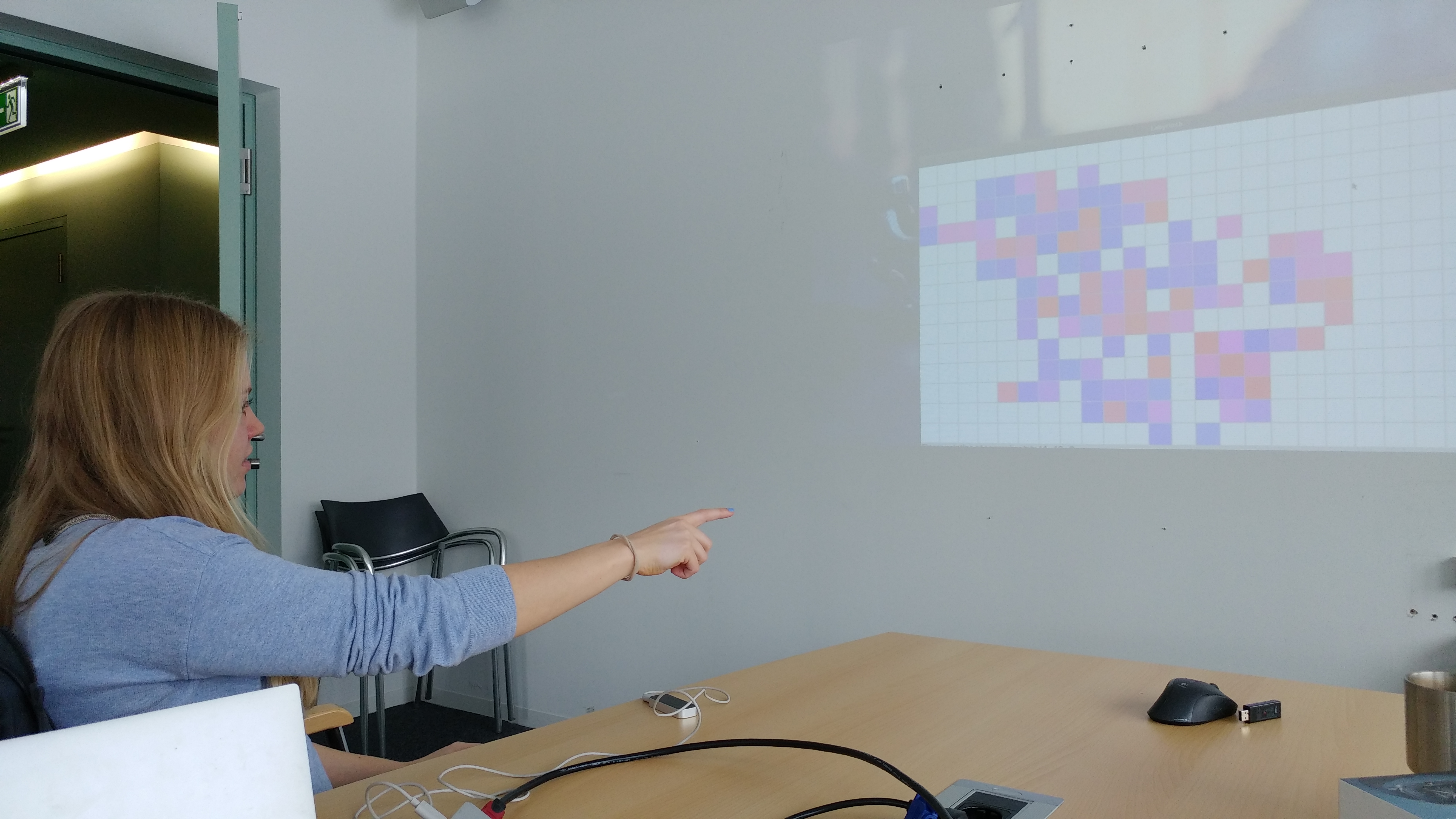}
    \caption{Setup and usage of the system, showing the projected display on a wall and a user who is mid-air pointing using the LeapMotion device to draw on the  projection.}
    \label{fig:usage-system}
\end{figure}

\subsection{User Interface}
\label{sec:UI}
The user interface is divided into two parts: 
a display, either shown on a monitor or projected to a nearby wall and a LeapMotion.

\subsubsection{Display}
The display contains a plain grid with panels of a fixed size. 
The colors of these panels can be changed by the user in cooperation with the learning algorithm. 
The grid has a size of 24 by 14 panels. 
At the beginning of the drawing application, all panels are white. 

While drawing, the Moore neighborhood around the user's finger is colored. Moore neighborhood describes the eight fields around the current finger position field. 
The colors are thereby selected by a learning algorithm, which learns the users color preferences.
Whenever the user moves from his current position to one of the neighbouring panels, this panel will keep its current color until the drawing is finished or the user points at the panel again.
The user can also change the proposed colors of the neighbouring panels by not moving for two seconds, if he does not like the current color selection. 
The intensity of the colors depends on the relative position of the finger to the tracking sensor. 
Moving the finger behind the controller leads to a higher opacity, moving in front to a lower one. 
Four different opacity options are available for each color.
This approach encourages the user to draw freely in three dimensional space in front of the screen. 
The color of each panel within the neighborhood is set using n-armed bandits, described in section \ref{sec:implementation}. 
For comparison during evaluation, a random mode can be set which chooses the colors of the Moore neighborhood randomly. This is used in the user study as control and test iteration. 
Figures \ref{fig:no-effect}, \ref{fig:effect}, and \ref{fig:med-effect} show examples of colored screens.

\subsubsection{LeapMotion}
The user's right hand is tracked using a LeapMotion controller.
This device is placed on a table about 30cm in front of the user, as shown in figure~\ref{fig:usage-system}. 
The device is connected to a computer, which processes the sensor data and executes the learning algorithm.
Ideally, the user should be able to point at all panels of the screen while keeping his fingers in the controller's field of view. 
This is necessary for an engaging drawing experience where loss of tracking or inability to reach some panels would be distracting.

Before starting the drawing application, a short calibration is necessary \cite{aslan2017pre}.
After placing the LeapMotion controller in front of the user, the user points to three panels in the different corners of the grid while being tracked by the controller.
The coordinates of the fingertip as measured by the LeapMotion and the grid-coordinates are used to calculate a transformation from LeapMotion detections to grid-coordinate space. 
After the calibration is finished, the drawing application can be started. 
Only one calibration is needed per experiment setup.

\subsection{Implementation}
\label{sec:implementation}
To process the LeapMotion inputs and realize the painting game or application with a personalized learning component, a Python application using PyGame for the UI was implemented. 
It uses an n-armed bandit algorithm, a vanilla reinforcement learning approach. 
It is called vanilla, because it doesn't use any frameworks or supporting libraries.
We choose to use this algorithm because the learning task requires no state transitions. 
The environment doesn't change its state within an action execution, so our initial approach to use a Q-learner, which selects actions depending on current state, would not have been useful.
Also, we need a fast visible learning progress, which provides quick feedback for the user. 
N-armed bandits are the best method in our case, because they immediately take the executed actions into account. Thereby we archive user personalized outputs up from the first iteration.

To calculate rewards for an action, the users' input is used. 
After each change of color, we need to wait until the user has either changed their position or remained in the same position for more than two seconds. 
Out of the before and after position a movement direction is calculated to reward the system.
Having to wait for the user's action slows down the learning process.
Since a bandit is not based on state transitions and produces only a very small action set, the algorithm is still able to learn fast and deliver immediately visible results. 
Based on evaluations of Vermorel et al. \cite{banditEmpiricalEvaluation}, who achieved good results at 1 000 iterations for 100 arms, we decided to do 500 iterations for 10 arms. Thereby we achieve an acceptable learning rate, in opposite to a neural network, which would require even in a very simple version more than 10 000 iterations. \cite{smith2018disciplined}

\begin{figure}[h]
    \vspace{-2pt}
    \centering
    \includegraphics[width=0.8\textwidth]{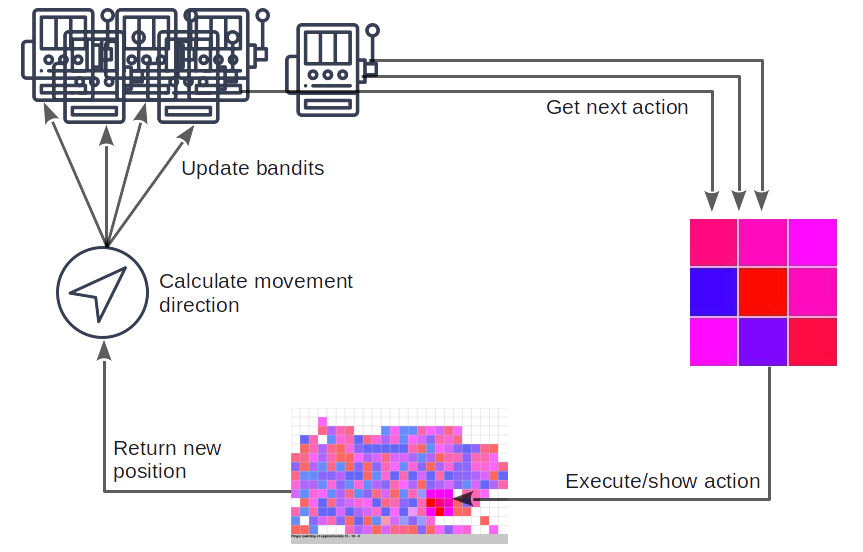}
    \caption{Learning process}
    \label{MAB}
    \vspace{-2pt}
\end{figure}

The agent consists of multiple bandits, each one being responsible for the color of one panel in the Moore neighborhood. 
A complete process overview can be seen in figure~\ref{MAB}.
In total, nine bandits are used to color the panels around the user's fingers. 
Each of those bandits has ten arms. 
The arms are mapped to predefined colors in the range of blue to red.
The arm or action selection is done using a greedy algorithm. 
It utilizes a value-function $Q_t(a)$, which computes the estimated value of action $a$ at time step $t$. 
To select the action $A_t$ the bandit uses the current action-value: 
\begin{equation}
A_t=argmax_a \ Q_t(a) 
\label{value-function}
\end{equation}
Thereby $argmax_a$ describes the value of $a$ at which the expression that follows
is maximized. The given action space is explored with an $\epsilon$-greedy method giving $\epsilon$ a value of 0.2 \cite{sutton2018reinforcement}.

The actions selected by the nine bandits are combined and displayed on the grid around the current finger position. 
As the user moves their finger, a new position is detected and is passed to the agent.
Out of the new and old positions the user's movement direction is calculated, which could be either left, right, up, down or stay.
This is then used as a basis for the bandits rewards.
The algorithm rewards only the bandits, which produced the outputs in the user's selected movement direction.
All other bandits do not get any reward.
Thereby, the bandits which produce the finger's von Neumann neighborhood receive a reward of value 1.0 and the diagonal fields of the Moore neighborhood receive a reward of 0.5. 
The action value function of the successful bandits are updated using an incremental update formula:
\begin{equation}
Q_t (a)=Q_{t-1} (a) + \frac{1}{N_t} (R_{t} - Q_{t-1}(a))
\label{value-function}
\end{equation}
Here, $a$ is the currently executed action, $R_{t}$ describes the received reward and $N_t$ describes the number of executed tries \cite{sutton2018reinforcement}. 
One experiment consists of 500 color changes, each followed by one user movement.

\section{User Study}
To explore how users would experience the use of the drawing application, we conducted a qualitative user study with 14 participants. While the participants used the drawing application, a researcher interviewed them by asking mainly open questions about their experience. 
The setup of the system during the user study can be seen in figure~\ref{fig:usage-system}.

\subsection{Participants and Procedure}
The user study was executed with 14 (10f, 4m) participants.
The studies participants have diverse working backgrounds out of multiple different domains.
Eight were aged between 17 and 25, five between 26 and 30 and one participant was of age 83.

After informing the participants about the overall setup and their interaction possibilities (i.e., to draw what ever they liked by pointing on the projected display), they were asked to fill in a demographic questionnaire, Additionally, we ask them about their experience with self-learning systems and what they thought of intelligent user interfaces in their daily lives.

Each participants started with receiving a short introduction into the study topic and exploring the technical setup with the LeapMotion.

When participants felt comfortable using the LeapMotion, they were asked to use the drawing application four times. In the first two sessions the system inked the eras poited by the system with random colors. Thus, in the first two sessions the system did not adapt to the user behavior at all.
We did this to allow the user to explore the drawing application and to reduce any bias in the session when they would be using an ``AI to draw with''. Thus, only in the last two drawing sessions, the multi-armed bandits were used to adapt the colors of the Moore neighborhood to each user's preferences. 

We encouraged the participants to talk about their feelings and thoughts while they were  drawing. 
As soon as they stared their third drawing and the system started to learn, we ask them if they noticed any differences to the previous drawing sessions. % 
After participants finished all drawing sessions, we asked them if  they preferred any of the drawing sessions most. Furthermore, we discussed with them their opinions on automated intelligence and if they have any suggestions for improving such system.

\subsection{Results} 
To show there was an adaption, we evaluated the data collected in the user study in three different ways. 
First, we identified general preferences based on the participants' statements. 
Second, we categorized participants into groups according to their self-assessed creativeness and technical knowledge and compared these groups. 
Third, we studied if system adaption was experienced by users.

\subsubsection{General user preferences}
Most participants (10 of the 14) stated that they preferred the 3th or 4th iteration, which are both adaptive versions. % the color suggestions to their movements. 
They named reasons like having a better understanding of the system, the generated colors, the nice figures they drew or just generally liking them more without being able to name a reason. But some also suggested that they had become accustomed to using mid-air pointing for drawing. Seven of them explicitly stated they enjoyed using the system and thought it was funny. 

However many of the participants missed an explicit goal of the task. Some participants asked what they should do, and whether they were using the system correctly or if there was any goal they should reach.

\subsubsection{Creativity and technical knowledge}

Before testing the drawing application, the participants were asked how they would self-describe their creativity and their technical background.
Based on their self assessment twelve of the 14 participants can be put into three main groups.
Six considered themselves creative having a non-technical background.
Three referred to themselves as a little creative having a non-technical background. Another three also felt a little bit creative, but had more technical knowledge.
One participant considered himself non-creative and reported that he had no technical knowledge and another person considered himself creative and he stated he had technical knowledge.
The last two participants will not be considered in the following discussion since their groups only consisted of one person each.

We will provide three participants in more detail to provide and in depth impression of participants' encounters with the drawing application. They are selected because they were so divers, and thus, their combined statements could be considered representative and inspiring. In order to reference them later on in the findings section, we gave them fictional names.
\\
\emph{Sportive Schoolgirl} is a teenage girl who likes to paint and does a lot of sports in her free time.
In her everyday life as a digital native, she is using technology but has no further interest in understanding it.
She is open to new technology and interaction techniques. 
\\
\emph{Computer Crack} is a male software engineer and interested in new technologies.
He called himself medium creative. 
In his work, he finds new, creative ways to solve software problems.
One part of it is to design algorithms.
\\
\emph{Poor Painter} is creative men who works in a callcenter to finance his art studies.
For his arts, he likes to try some new technologies.
He believes that you could use a LeapMotion controller as a virtual reality drawing tool.

Within the creative - non-technical group, five of six participants thought that the interaction was funny and interesting, but only three would use the LeapMotion or adaptive systems in their day-to-day lives.
They were either focused on drawing shapes or drawing in one color.
Two concentrated only on colors, two on figures and two on both. 
For five of them, the system adapted to a certain group of colors. 
Half of them liked the iterations with adaption the best, the others had no preferences.

\emph{Sportive Schoolgirl} and \emph{Poor Painter} are both part of this group.
During the second adaptive iteration \emph{Sportive Schoolgirl} noticed \emph{'how funny it is, everything gets blue'}.
She liked this way of drawing and focusing on colors. 
She also noted that she would like to use the system to design mosaic tiles for her bathroom.

\emph{Poor Painter} would have liked to be able to choose certain colors for drawing during the random iterations. 
Still, he liked using the drawing application and stated it was \emph{'really funny'}. 
He suggested using the system to get out of a creative block by drawing randomly at first. 
Also he had an interesting usage idea: \begin{quote}
    After a while, one would start seeing shapes in the colored tiles and could start filling them in. 
This method could change your way of creative thinking. 
\end{quote}

The second group was slightly creative and had no technical knowledge.
All three participants were focusing on colors and one has an additional focus on drawing figures.
Even with this focus, for one participant the system showed no adaption to a certain color. 
Still, this participant noticed a difference when playing the adaptive iterations. 
For the other users, a slight adaption of color can be seen, which was not noticed by the users.

All three members of this group preferred the second learning iteration and would use the LeapMotion and adaptive systems again. 
Since the system only showed little or no adaption in color, their preference was most likely based on their improved understanding of the interaction with the sensor.

The third group consists of three slightly creative participants with some technical knowledge.
Two of them had a focus on colors. They both liked the drawing application and thought it was \emph{'funny'}.
The third one had no noticeable focus on any specific part of the system. 
Even without paying attention to the colors, some adaption of color can be seen for this user, which most likely happened completely unconsciously. 
All three users liked the reinforcement learning iterations most and would use systems with automated intelligence and the LeapMotion again.

\emph{Computer Crack} is part of this group. He noticed a feeling of disconnectedness between the screen and his hand, feeling like his movements were not translated perfectly to the grid. 
Despite his difficulties with using the LeapMotion, he also pointed out \emph{'how cool he thought the drawing application was'}.

\subsubsection{Adjustment}

\begin{figure}[h]
    \centering
        \begin{subfigure}{0.4\textwidth}
            \includegraphics[width=\textwidth]{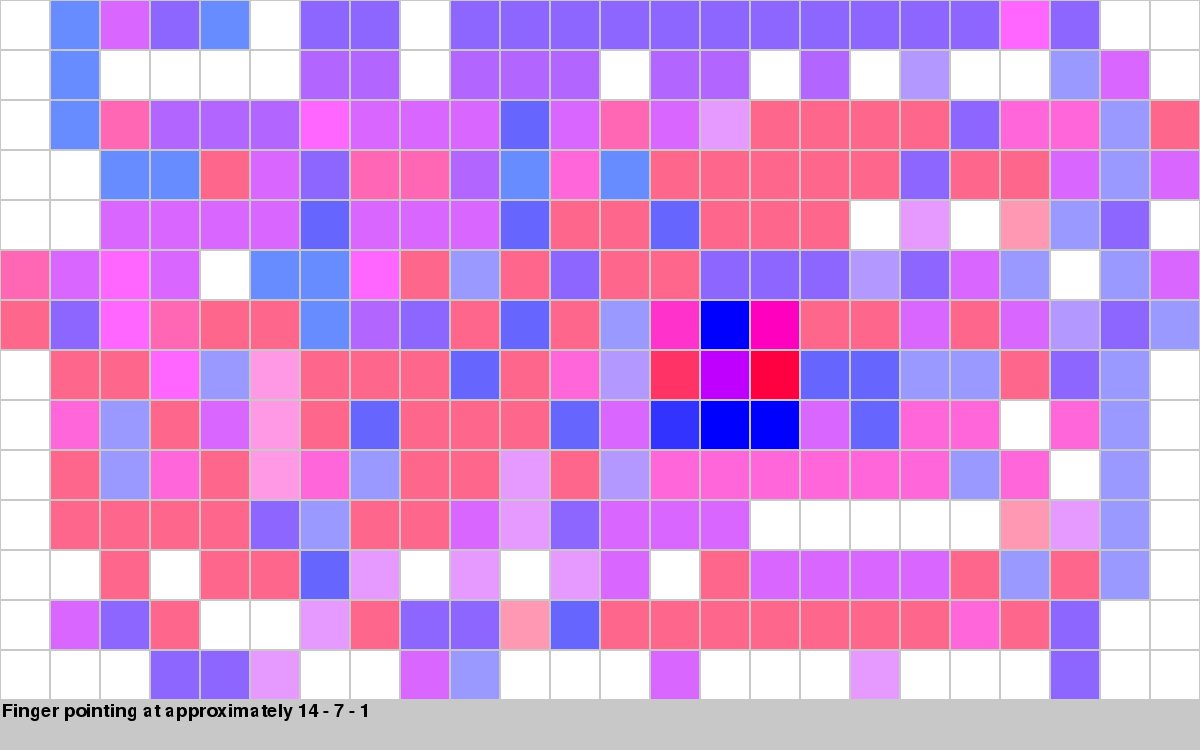}
            \caption{random coloring - it. 1}
            \label{fig:no-effect:r1}
        \end{subfigure}
        \begin{subfigure}{0.4\textwidth}
            \includegraphics[width=\textwidth]{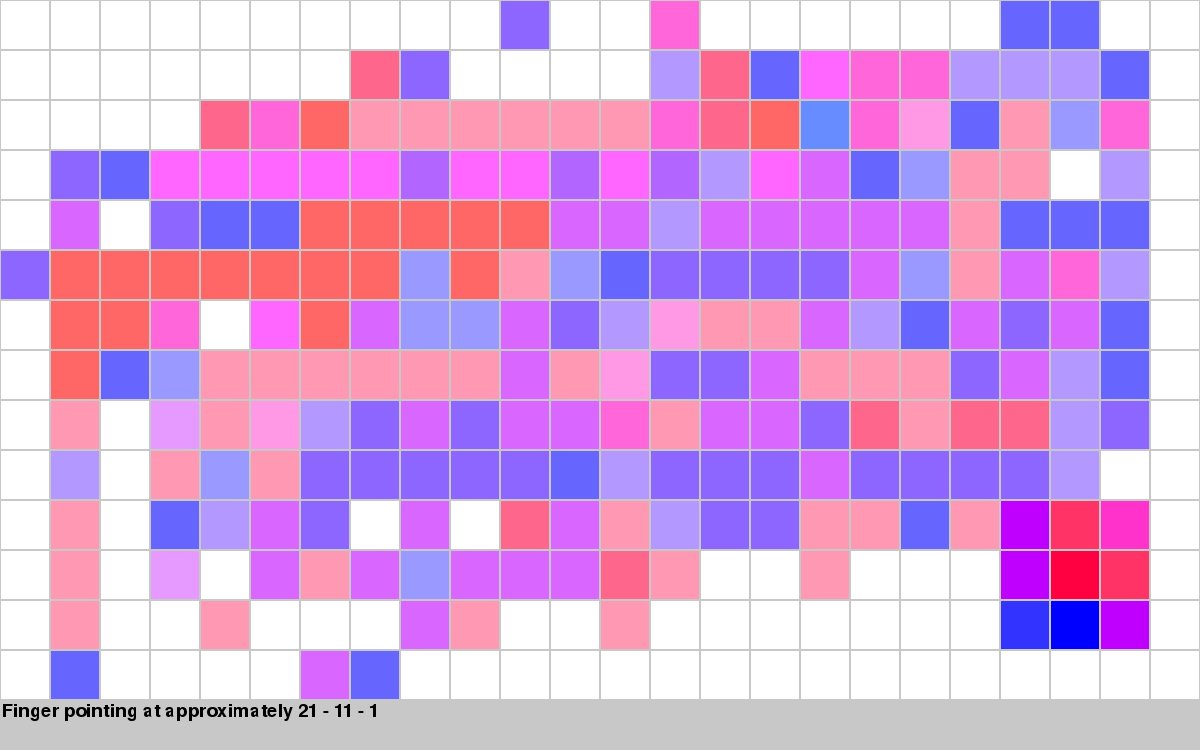}
            \caption{random coloring - it. 2}
            \label{fig:no-effect:r2}
        \end{subfigure}
        \begin{subfigure}{0.4\textwidth}
            \includegraphics[width=\textwidth]{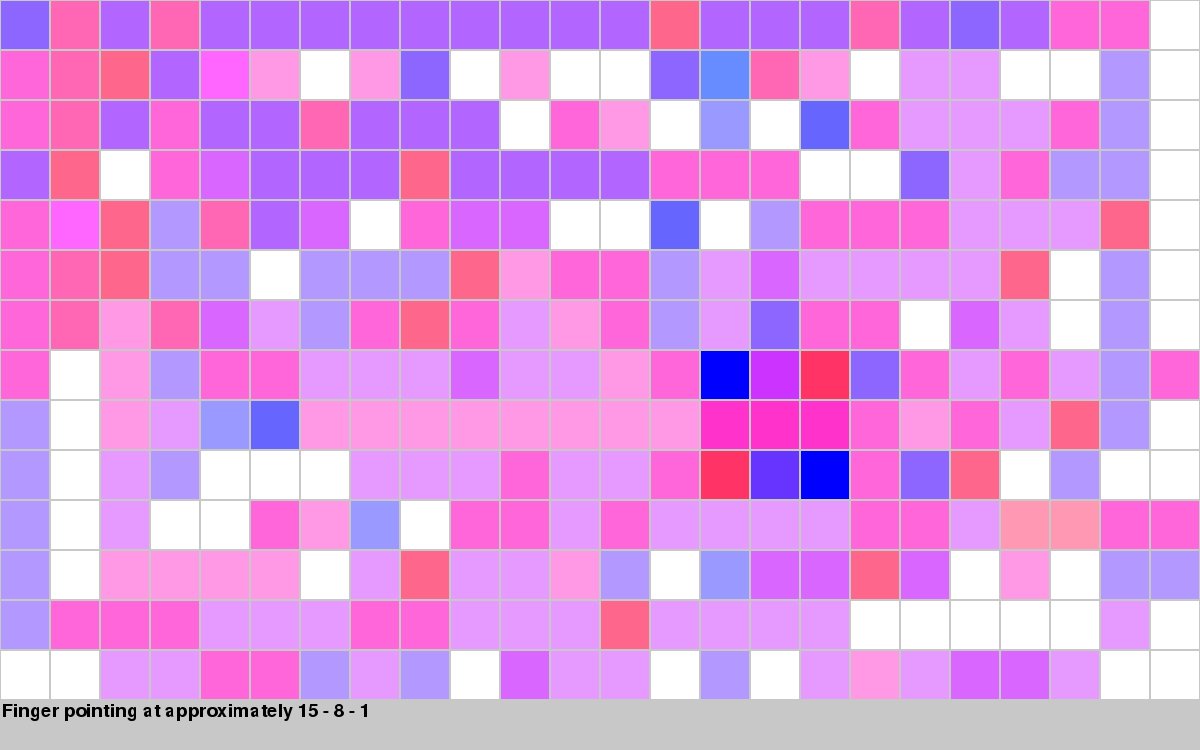}
            \caption{trained coloring - it. 1}
            \label{fig:no-effect:t1}
        \end{subfigure}
        \begin{subfigure}{0.4\textwidth}
            \includegraphics[width=\textwidth]{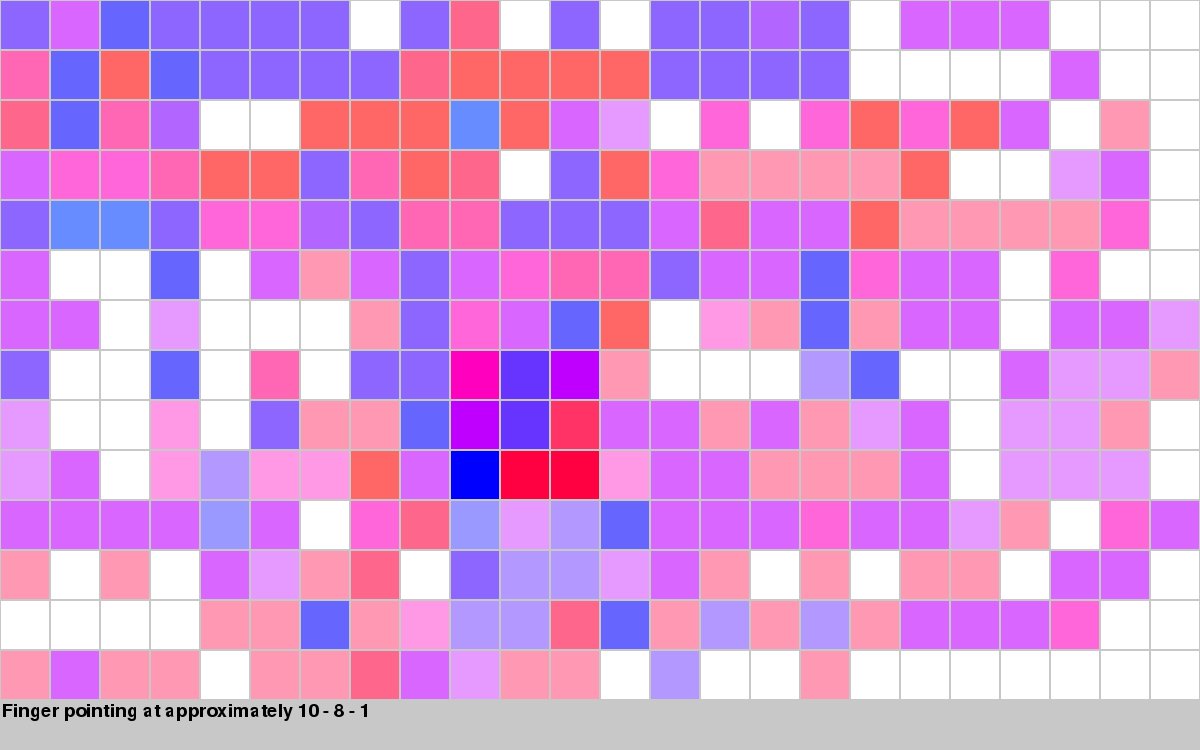}
            \caption{trained coloring - it. 2}
            \label{fig:no-effect:t2}
        \end{subfigure}
    \caption{Examples of the screen with untrained/random and trained coloring from a user with focus on structures and figures and no training effect.}
    \label{fig:no-effect}
\end{figure}

We examined the drawings of each participant during the random and the learning iterations. 
For nine participants, the system ended up suggesting mainly colors from one color group. 
Some tendency towards a certain color group can be seen for four other participants. 
Only the drawings of one participant showed no adjustment compared to the drawings with random color suggestions. The drawings of this participant can be seen in figure~\ref{fig:no-effect}.

\begin{figure}[h]
    \centering
    \begin{subfigure}{0.4\textwidth}
        \includegraphics[width=\textwidth]{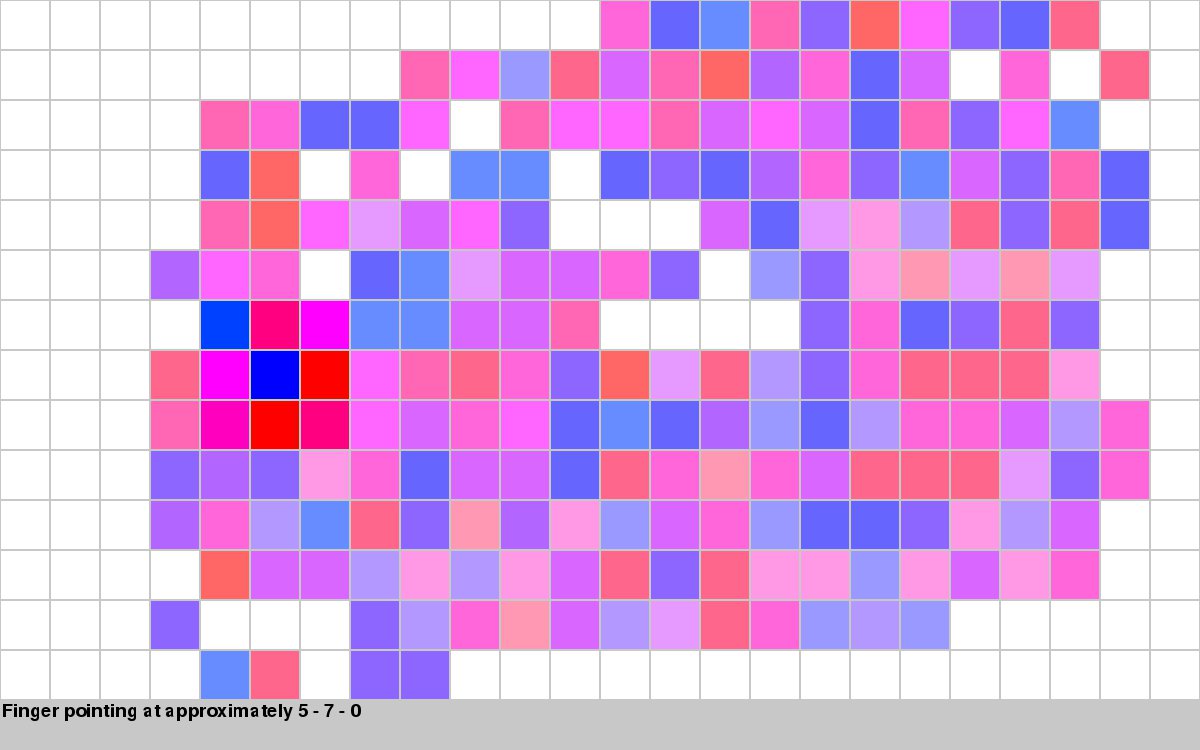}
        \caption{random coloring}
        \label{fig:effect:random}
    \end{subfigure}
    \begin{subfigure}{0.4\textwidth}
        \includegraphics[width=\textwidth]{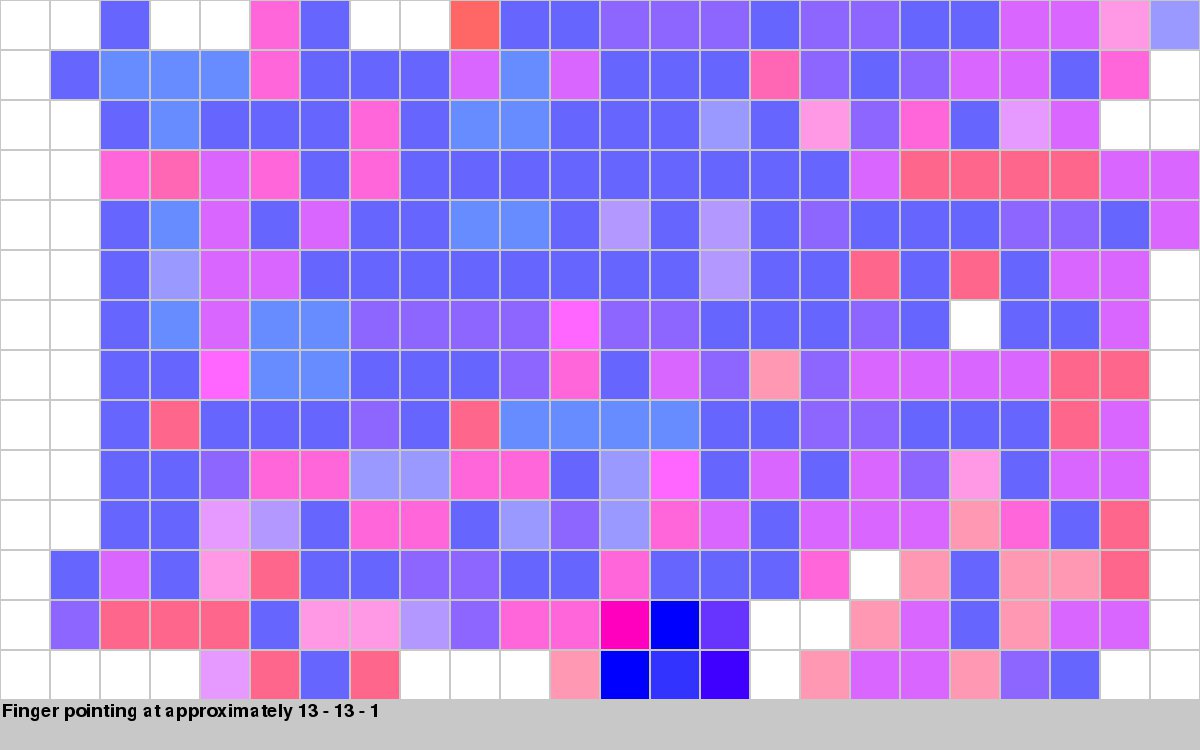}
        \caption{trained coloring}
        \label{fig:effect:training}
    \end{subfigure}
    \caption{Examples of the screen with untrained/random and trained coloring from a user with focus on colors and a positive training effect.}
    \label{fig:effect}
\end{figure}

Five out of the nine participants whose pictures showed a clear focus towards one color group said the system was funny. 
Five respectively could see themselves using adaptive systems or sensors like the LeapMotion in their everyday lives. 
Six of them preferred the adapting iterations. 
Four participants focused on the colors, four mainly concentrated on drawing shapes. 
An example of the color adaption can be seen in figure \ref{fig:effect}. 

\begin{figure}[h]
    \centering
        \begin{subfigure}{0.4\textwidth}
            \centering
            \includegraphics[width=\textwidth]{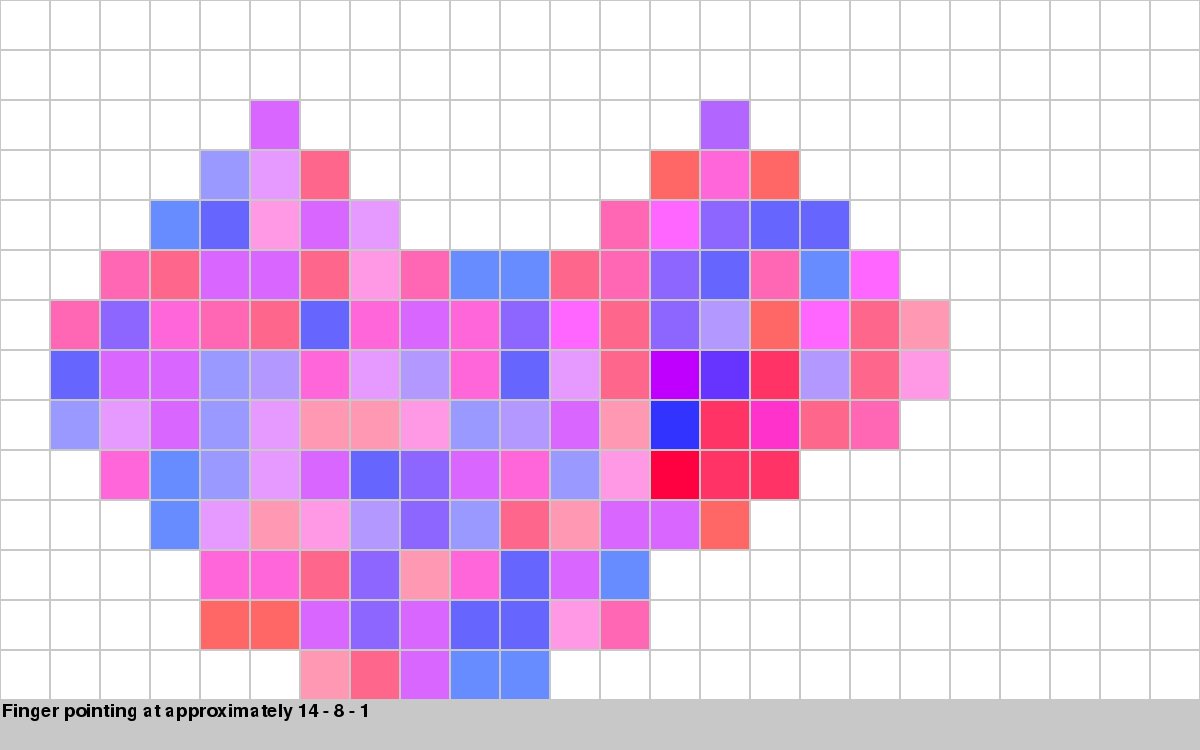}
            \caption{random coloring}
            \label{fig:med-effect:random}
        \end{subfigure}
        \begin{subfigure}{0.4\textwidth}
            \centering
            \includegraphics[width=\textwidth]{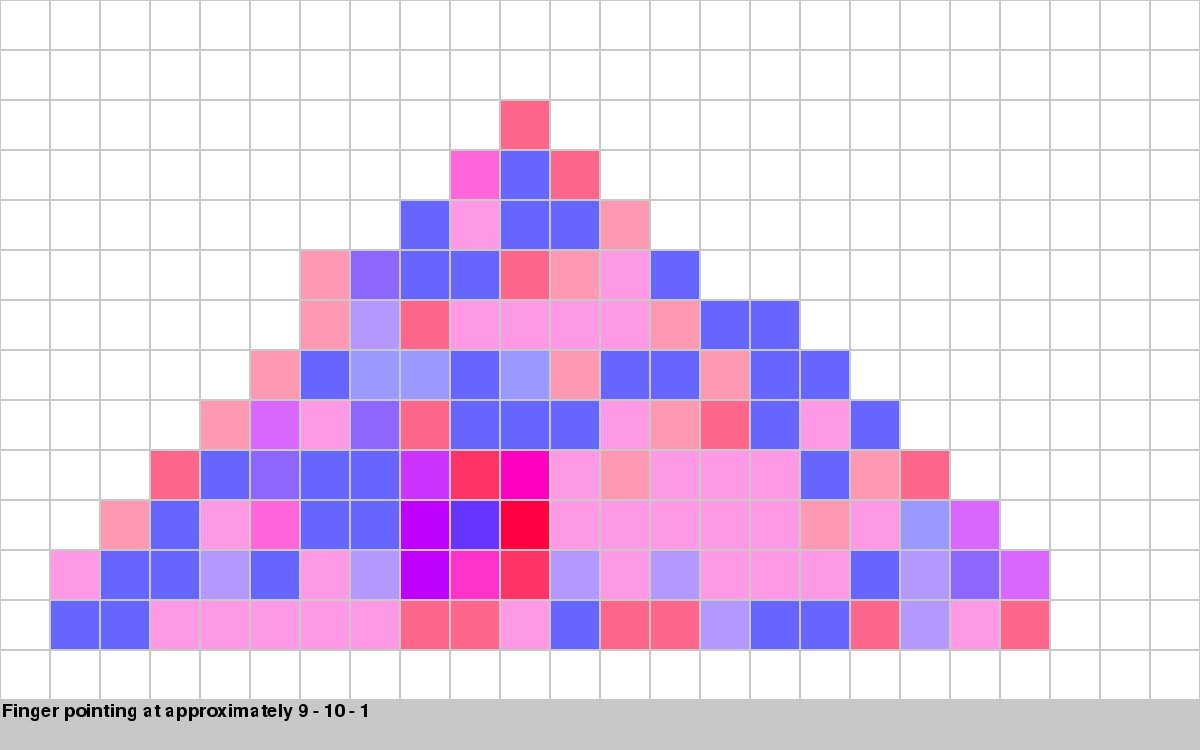}
            \caption{trained coloring}
            \label{fig:med-effect:training}
        \end{subfigure}
    \caption{Examples of the screen with untrained/random and trained coloring from a user with focus on structures and figures and no really training effect.}
    \label{fig:med-effect}
\end{figure}

Out of the four participants with slight color adaption, three would like to keep using adaptive systems and all of them could see themselves using sensors like the LeapMotion in the future. 
Three preferred playing the adaptive iterations, one user had no preferences. 
This group contains users who focused on the colors, on drawing figures or on understanding the system. 
Despite their differences in using the system, a tendency towards certain colors can be seen in their drawings, even when they only focused on drawing shapes. 
Two drawings of a user from this group with a focus on drawing shapes can be seen in figure \ref{fig:med-effect}.

\section{Conclusion and Future Work}
In this paper, we first described in detail a prototype of an intelligent digital canvas, which is able to learn from a user's real-time mid-air pointing behavior, their preferences in coloring specific ``pixels'' or panels on the canvas. 
We then reported on results of a qualitative study with a heterogeneous group of participants. 
Our overall aim was to explore how people perceive drawing ``with an AI'', or more specifically how they would feel to draw on a canvas, which learns to choose drawing colors autonomously based on the users' previous tendencies.   
Overall, participants were intrigued and engaged by the drawing application. Participants who consider themselves less creative seem to be more willing to draw something together with an AI, which could be because they may benefit more from using an (semi)autonomous tool.   
Despite the user study revealing some limitations, such as not exploring variations of the learning algorithm, we believe that the qualitative nature of the study represents a good starting point. And, results hopefully will also inspire fellow researchers to apply reinforcement learning to further the study of drawing-related interactions with smart pervasive displays. %michelle: multi-clause sentence

In our future work, we aim to study how multiple users could collaborate on a drawing, with a system learning over longer periods of time, for example, from by-passers at a public space with each by-passer potentially contributing to a drawing.

\bibliographystyle{unsrt}
\bibliography{sources}

\end{document}